\documentclass[pra,onecolumn,tightenlines,showpacs,amsmath,amssymb,nofootinbib]{revtex4}
\usepackage{amssymb}
\usepackage{amsmath}
\usepackage{colordvi}
\usepackage{amsthm}
\usepackage{epsfig}

\begin{document}

\newtheorem{theorem}{Theorem}
\newtheorem{lemma}{Lemma}
\newtheorem{corollary}{Corollary}

\title{Unitary transformations can be distinguished locally}
\author{Xiang-Fa Zhou}
\email{xfzhou@mail.ustc.edu.cn}
\author{Yong-Sheng Zhang}
\email{yshzhang@ustc.edu.cn}
\author{Guang-Can Guo}
\affiliation{\textit{Key Laboratory of Quantum Information,
University of Science and Technology of China, Hefei, Anhui
230026, People's Republic of China}}

\begin{abstract}
We show that in principle, $N$-partite unitary transformations can
be perfectly discriminated under local measurement and classical
communication (LOCC) despite of their nonlocal properties. Based
on this result, some related topics, including the construction of
the appropriate quantum circuit together with the extension to
general completely positive trace preserving operations, are
discussed.
\end{abstract}

\maketitle

Superposition plays the central role in quantum mechanics. The
quantum nonorthogonality and entanglement due to superposition,
which show many counter-intuitive behaviors compared with those in
classical world, have drawn much attention in the past two
decades. Quantum nonorthogonality put many constraints on
physically accessible manipulations on input states. It is
well-known that two nonorthogonal pure state can not be perfectly
discriminated \cite{discrimination}. On the other hand, quantum
nonlocality due to entanglement, which was first brought into
attention by Einstein, Podolsky, and Rosen (EPR) in 1935
\cite{EPR}, is also one of the most interesting and important
parts in quantum information science. Today, quantum entanglement
has been viewed as a significant resource for quantum information
processing, and currently the behavior of entanglement in quantum
information science is still under investigation.

Although perfect identification of nonorthogonal quantum states
are impossible in quantum world, when we refer to quantum
operations, thing becomes very different. It was proved that two
unitary operations can be perfectly discriminated after applying
the unitary gate a finite number of times in parallel
\cite{unitary,unitary2}. On the other hand, the nonlocality of
unitary transformation has been extensively studied because of its
fundamental importance during the construction of universal
quantum circuit \cite{gates}. For example, it has been shown that
a sequence of a nonlocal gate (e.g., Control-not gate or
Control-phase gate) and single-qubit rotations can be used to
construct any desired transformations. Also nonlocal gate can be
classified and simulate each other under specific conditions
\cite{nonlocal1,nonlocal2}. Based on these results, one natural
problem arises - what is the influence of the nonlocality of
quantum operation on the discrimination.

In this work, we consider to discriminate two unitary
transformations with local methods. Compared with its counterpart,
i.e., local identification of quantum states, which is often
considered for orthogonal states \cite{locc1,locc2}, we find that
any two unitary transformations can be perfectly identified
locally despite of their nonlocal properties.

Before concentrating on the specific topics, let us make a few
remarks about the difference between the discrimination of quantum
states and of quantum operations. Generally to identify a quantum
state, one should make a measurement on the given state followed
by an estimation based on the measurement results. Such process
usually collapses the input states which thus cannot be used any
more. However, thing becomes different when we refer to quantum
operations. The reason lies in the fact that quantum operations
never collapse, and in principle it can be repeated any times if
we need. What's more, when unitary operations are considered, by
exchanging the input and output ports of the whole setup, we can
obtain the reverse transformations. Actually, these facts make the
discrimination of quantum operations very different from that of
quantum states.

Generally the strategy of operation identification is formulated
as this: we employ a quantum circuit $f(U)$ which is made up of
the selected operation $U$ on the suitable input state
$\rho_{s,a}$, where $s(a)$ denotes the circuit system (auxiliary
system). If only local methods are required, $\rho_{s,a}$ must
also be separable. To obtain the maximal distinguishability, the
overlap of the output states should be as small as possible for
different quantum operations. Fig. 1 shows the sketch of the
identification process under local operation and classical
communication (LOCC). When global operation are permitted, both of
the circuit and the input state can be constructed to realize a
perfect discrimination for unitary transformations
\cite{unitary,unitary2}. However, when only local operations and
resources are permitted, thing becomes not so obvious. To simplify
our consideration, in the following, we mainly focus on bipartite
system.

\begin{figure}[ht]
\mbox{\epsfig{file=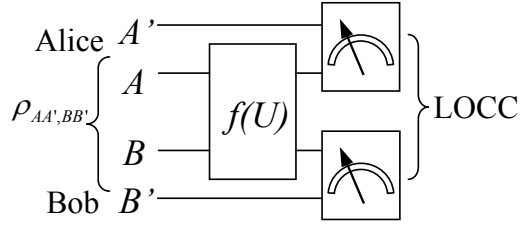,width=7.cm}} \caption{Illustration of
the identification of unitary transformations under local
operation and classical communication. Alice and Bob input a
locally implemented state $\rho_{AA',BB'}$ to a quantum circuit
$f(U)$ followed by local measurement operations. The measurement
results are transmitted through classical channels to realize
perfect discrimination.}
\end{figure}

Let us begin with some simple observations. Here we mainly
concentrate on unitary operations, one can check that some of the
discussions are also suitable for general quantum operations. As
we have mentioned above, to realize perfect identification, one
need to find a suitable input state such that the corresponding
output states are orthogonal to each other for different selected
operations. Assume that we want to discriminate two unitary
operations $U$ and $V$. By inputting a locally implemented quantum
state $\rho_{AA',BB'}=\sum_i \lambda_i \rho^i_{AA'} \otimes
\rho^i_{BB'}$, we have that the two output states $\rho_U=(U
\otimes I_{A'B'}) \rho_{AA',BB'} (U^{\dag} \otimes I_{A'B'})$ and
$\rho_V=(V \otimes I_{A'B'}) \rho_{AA',BB'} (V^{\dag} \otimes
I_{A'B'})$ should be orthogonal to each other. Now consider the
spectral decompositions of $\rho^i_{AA'} = \sum_j
r_j|r^i_j\rangle_{(AA')}\langle r^i_j|$ and $\rho^i_{BB'}= \sum_k
s_k|s^i_k\rangle_{(BB')}\langle s^i_k|$. The requirement of
$\rho_U \perp \rho_v$ is equivalent to $(U \otimes I_{A'B'})
|r^i_j\rangle_{AA'}|s^i_k\rangle_{BB'} \perp (V \otimes I_{A'B'})
|r^{i'}_{j'}\rangle_{AA'}|s^{i'}_{k'}\rangle_{BB'}$ for any $i$,
$i'$, $j$, $j'$, $k$, $k'$. This observation shows in general, a
pure input state $|r\rangle_{AA'}|s\rangle_{BB'}$ is enough to
perfectly discriminate two unitary operations if they can.
Moreover, since two orthogonal pure states can be locally
identified \cite{locc1,locc2}, hence in this case $U$ and $V$ can
also be discriminated with local methods.

Consider two unitary transformations $U_{AB}$ and $V_{AB}$ with
zero overlap in trace norm, i.e.,
$\mbox{Tr}(V_{AB}^{\dag}U_{AB})=0$. Then by preparing the
following locally maximal entangled state as the input
\begin{eqnarray} \label{entanglement}
| \phi \rangle_{AB,A'B'}=|\phi\rangle_{AA'} \otimes
|\phi\rangle_{BB'},
\end{eqnarray}
where $|\phi\rangle_{AA'}=\sum_i|i\rangle_A|i'\rangle_{A'}$ (or
$|\phi\rangle_{BB'}=\sum_i|i\rangle_B|i'\rangle_{B'}$) is a
nonnormalized entangled state between the system $A$ and the
corresponding local environment $A'$ (or $B$ and $B'$). From the
following equation
\begin{eqnarray} \label{norm}
\langle \phi | V_{AB}^{\dag}U_{AB} \otimes I | \phi \rangle =
\mbox{Tr}(V_{AB}^{\dag}U_{AB})=0,
\end{eqnarray}
one immediately obtain that the two output states $U_{AB} \otimes
I |\phi\rangle_{AB,A'B'}$ and $V_{AB} \otimes I
|\phi\rangle_{AB,A'B'}$ are orthogonal to each other, hence can be
locally discriminated perfectly. Equations ($\ref{entanglement}$,
$\ref{norm}$) can be viewed as the extension of Jamiolkowski
isomorphism in local case \cite{nonlocal1}. The input state $|
\phi \rangle_{AB,A'B'}$ is universal for any two operations $U$
and $V$ satisfying $\mbox{Tr}(V^{\dag}U)=0$. Actually, given $U$
and $V$, if global input states are permitted, one can always
choose a suitable pure input state in the composite system of only
$A$ and $B$, namely, the auxiliary system can be neglected in this
case \cite{unitary,unitary2}. However, if only local resources are
required, in order to achieve the maximal distinguishability of
the output states, an entangled state between the system and the
environment seems to be required unless the global optimal pure
state is separable.

In the above case, perfect identification can be realized in a
single run for both global and local methods. In the more general
cases, one needs to run the selected gate $N$ times ($N$ is
finite). The optimal $N$ has been found for global discrimination
of $U$ and $V$, which asserts that if the minimal arclength
$\delta$ spread by the eigenvalue of $(U^{\dag}V)^{\otimes N}$ in
the circle $|z|=1$ is not less than $\pi$, then a perfect
discrimination scheme is allowed. Now assume $U^{\dag}V=U_1
\otimes U_2$ to be local operation, with $\delta_1$ and $\delta_2$
being the minimal arclengths of $U_1^{\otimes N}$ and
$U_2^{\otimes N}$ respectively. Then perfect global discrimination
can be implemented by inputting an entangled state if
$\delta_1+\delta_2 \ge \pi$. However, if only local input states
(e.g., $|r\rangle|s\rangle$) are allowed, since
\begin{eqnarray}
&&\langle r|U_1^{\otimes N}|r \rangle \langle s|U_2^{\otimes N}|s
\rangle =0 \nonumber \\
&\Leftrightarrow& \langle r|U_1^{\otimes N}|r \rangle =0 \mbox{ or
} \langle s|U_2^{\otimes N}|s \rangle =0,
\end{eqnarray}
this indicates that to distinguish $U$ and $V$ locally, at least
one of the two arclength $\delta_1$ and $\delta_2$ must be not
less than $\pi$. Therefore, generally in the local case the
optimal running times $N$ of the selected operation should be
greater than that of the global case.

As a special example, consider the following control unitary
transformation $U^{\dag}V=P_1 \otimes I + P_2 \otimes u$, where
$P_iP_j=\delta_{ij}P_i$ and $\sum_i P_i =I$, $I$ is the identity
operation, and $u$ is a local unitary manipulations. The
eigenvalues $r_i$ of $U^{\dag}V$ belong to the set $\{1, b_1, b_2,
...\}$ with $b_i$ and $|b_i\rangle$ being the eigenvalues and
eigenvectors of $u^{\otimes N}$ separately. If only local input
state $\rho_A \otimes \rho_B= \mbox{Tr} (
|\psi_{AA'}\rangle\langle \psi_{AA'}|\otimes |\psi_{BB'}\rangle
\langle \psi_{BB'}| )$ is permitted, then
\begin{eqnarray} \label{C-U2}
\mbox{Tr}[(U^{\dag}V)^{\otimes N}(\rho_A\otimes
\rho_B)]=x+(1-x)\sum_i b_i\langle b_i|\rho_B|b_i\rangle,
\end{eqnarray}
where $x=\mbox{Tr}(P_1\rho_A) \ge 0$ and can be chosen arbitrarily
by input appropriate $\rho_A$. In order to make the
right-hand-side of Eq. ($\ref{C-U2}$) to be zero, one can easily
obtain that the minimal angular spread of $\{1, b_1, b_2, ...\}$
should be not less than $\pi$. Therefore, in this case the minimal
$N$ required equals to that of global case. Similarly, suppose
$U^{\dag}V=(P_1 \otimes I + P_2 \otimes u) \cdot (U_1 \otimes
U_2)$ and $U2 \neq I$ or $U2 \neq u^{\dag}$. If $U2 \neq
u^{\dag}$, then by inputting appropriate state $|\psi\rangle_A
|\psi\rangle_B$ with $|\psi\rangle_A$ lying in the support of
$P_1$, $U^{\dag}V$ is equivalent to the local transformation
$(uU_2)|\psi\rangle_B$, hence can be perfectly identified.

In the above discussions, we have considered to discriminate
several special kinds of unitary transformations. They all can be
perfectly identified and the optimal quantum circuit and input
state can be easily obtained. In the following, we mainly focus on
the most general case. Although we cannot present the optimal
quantum circuit and input state, we prove that, in principle, any
two unitary operations $U$ and $V$ can be perfectly identified
locally.

Following \cite{exactuniverse}, we call a $2$-qudit gate $U_{AB}$
to be primitive if $U_{AB}$ maps a separable state to another
separable state; otherwise, $U_{AB}$ is imprimitive. Generally, a
primitive gate $U_{AB}$ can be expressed as the product of 1-qudit
gate up to a swap operation $P$, namely, $U_{AB}=U_A \otimes U_B$
or $U_{AB}=U_A \otimes U_B \cdot P$ with $P|\alpha \rangle_A
|\beta\rangle_B=|\beta \rangle_A |\alpha \rangle_B$. For
simplicity, in the following, we use $H$ to denote the set of all
2-qubit gates of the form $U_A \otimes U_B$. Under these
assumptions, we then introduce the following lemma.
\begin{lemma}
$H$ together with an imprimitive gate $Q$ can generate the unitary
group $U(d^2)$.
\end{lemma}

A detailed proof of this lemma can be found in
\cite{exactuniverse}, which is used to study the university of
quantum gate. This lemma indicate that if $Q$ and all local
unitary transformations are permitted, we can then construct
$H'=QHQ^{-1}$. By choosing suitable sequence of $H$ and $H'$, we
can obtain any desired elements in $U(d^2)$. The length of the
sequence is finite, therefore it is only need to run the
imprimitive gate a finite number of times.

Based on this lemma, we now prove the main theorem of this work.

\begin{theorem}
Any two unitary transformation $U_{AB}$ and $V_{AB}$ can be
perfectly identified with local methods.
\end{theorem}

Proof: Following our former discussions, we obtain that if both
$U_{AB}$ and $V_{AB}$ are primitive, then they can be perfectly
discriminated locally.

Now assume that only one of the two unitary gate is primitive.
Without loss of generality, we suppose $V_{AB}$ to be imprimitive.
According to the lemma, we obtain that there exists a quantum
circuit $f(V_{AB})$ made up of the elements in $H$ and
$H'=V_{AB}HV^{\dag}_{AB}$ such that $f(V_{AB}) \in (HH')^n$ is
some control unitary transformation. On the other hand, since
$U_{AB}$ is primitive, which means $H'=U_{AB}HU^{\dag}_{AB}=H$,
one immediately obtain that $f(U_{AB})$ is also primitive. Because
$f(U_{AB}) \neq f(V_{AB})$, we have that the two unitary
operations can be locally identified.

If $U_{AB}$ and $V_{AB}$ are both imprimitive, Following the
lemma, we obtain that there is a quantum circuit such that
$f(U_{AB})=e^{i L^A_{12} \otimes L^B_{12}}$ with
$(L^A_{12})_{ij}=\delta_{i1}\delta_{j2}+\delta_{i2}\delta_{j1}$
(or $(L^B_{12})_{ij}$). If $f(V_{AB})$ is primitive, then perfect
local discrimination can be realized. Otherwise, both $f(U_{AB})$
and $f(V_{AB})$ are imprimitive. Since
$f(U_{AB})^{\dag}=A.f(U_{AB}).A^{\dag}$ with
$A=\mbox{diag}\{\sigma_z,I_{(d-2)}\} \otimes I$, $I \otimes
\mbox{diag}\{\sigma_z,I_{(d-2)}\}$, $\mbox{diag}\{\sigma_y,
I_{(d-2)}\} \otimes I$, or $I \otimes
\mbox{diag}\{\sigma_y,I_{(d-2)}\}$. One can easily check that if
the similar result occurs for $V_{AB}$, then $f(V_{AB})$ can be
expressed as $f(V_{AB})=e^{i x L^A_{12} \otimes L^B_{12}}$ for
some $x \in \mathbb{R}$. Therefore the whole question can be
divided into the following two parts:

\emph{i}). If $f(V_{AB})\neq e^{i x L^A_{12} \otimes L^B_{12}}$
for any $x \in \mathbb{R}$, then by employing the transformation
$Af(\cdot)A^{\dag}f(\cdot)$, we can obtain an identity operation
for $U_{AB}$. Because $Af(V_{AB})A^{\dag }f(V_{AB}) \neq I$, the
two operations thus are locally distinguishable.

\emph{ii}). If $f(V_{AB})=e^{i x L^A_{12} \otimes L^B_{12}}$, then
when $x \neq 1$, $f(U_{AB})$ and $f(V_{AB})$ can be reduced to
$e^{i L^A_{12}} \otimes I$ and $e^{ix L^A_{12}} \otimes I$ by
inputting a product state $|\phi\rangle |\psi\rangle$ with
$|\psi\rangle$ being an eigenvector of $L^A_{12}$, which,
therefore, can be perfectly identified locally by running the
circuit a finite number of times in parallel. Otherwise we have
$f(U_{AB})=f(V_{AB})$. Since $e^{i L^A_{12} \otimes L^B_{12} }$ is
imprimitive, it can be used to construct the desired operator
$U^{\dag}_{AB}$. Thus the original problem is reduced to the
locally identification of the identity operation and
$U^{\dag}_{AB}V_{AB}$, which can be implemented perfectly.

This completes the proof.

The above theorem shows that in principle, to realize a perfect
local identification, we only need to run the selected unitary
operation a finite number of times. Although we have assumed that
the two subsystems $A$ and $B$ have equal dimensions, one can
easily obtain that the same result holds even if $A$ and $B$ have
different dimensions. For example, if $dim{{\cal H}_A} < dim{{\cal
H}_B}$, then by introducing another subsystem $A_1$ in Alice's
side such that $dim{{\cal H}_A} + dim{{\cal H}_{A_1}} = dim{{\cal
H}_B}$, we can obtain two extended unitary transformations $U
\oplus I_{A_1}$ and $V \oplus I_{A_1}$, which thus can be
identified with the methods described above.

It should be mentioned that the ancillary subsystem $A_1$ usually
plays nontrivial role during the discussion of operation
discrimination \cite{duanprivite}. In practice, given two
different operations $\{\xi_1, \xi_2 \}$ acting on the same
Hilbert space $A$, it is always possible to prepare a larger
system $A'$ such that $A'=A\oplus A_1$. Therefore, the original
problem can be reduced to the discrimination of the two newly
defined operations $\{\xi_1\oplus I_{A_1},\xi_2\oplus I_{A_1}\}$.
For instance, in the global discrimination of two unitary
operations $\{U, V \}$, the minimal running times usually reads
$N=\left[ \frac{\pi}{\delta} \right]$. However, when subsystem
$A_1$ is concerned, if $1$ is not one of the eigenvalues of
$(U^{\dag}V)^{\otimes N}$, and the two minimal arclengthes
$\{\delta, \delta' \}$, spread by the eigenvalues of
$(U^{\dag}V)^{\otimes N}$ and $(U^{\dag}V\oplus I_{A_1})^{\otimes
N}$ separately, are different, then we have $N' = \left[
\frac{\pi}{\delta'} \right] \ge N$. The subsystem $A_1$ can be
used to distinguish two unitary operations up to a phase factor.
For example, consider a three level system $\{ |0\rangle,
|1\rangle, |2\rangle \}$. Suppose the Hamiltonian of the whole
system is $H= \omega (|0\rangle\langle 0| + |1\rangle\langle 1|)$.
If we are restricted in the subspace $\{|0\rangle, |1\rangle \}$,
then when $T= \pi/\omega$, we obtain $U=-\mbox{diag}\{1,1\}$,
which cannot be discriminated from the identity operators $I$.
However, if the ancillary level $|2\rangle$ is concerned, then
perfect identification can be implemented by preparing suitable
pure input state in the total Hilbert space.

From the practical viewpoint, it will be valuable if one can
provide an optimal circuit to implement such kind of
identification operation \cite{exact-example}. Generally, it is
not easy to do this. Here, to simplify our consideration, we take
two-qubit gates as an example.

For any two-qubit unitary transformation $U$, it has the following
canonical decomposition \cite{two-qubit}
\begin{eqnarray} \label{decomposition}
U=(U_1 \otimes U_2) e^{i(h_x \sigma_x \otimes \sigma_x + h_y
\sigma_y \otimes \sigma_y + h_z \sigma_z \otimes \sigma_z )} (U_3
\otimes U_4),
\end{eqnarray}
where $\sigma_x$, $\sigma_y$, $\sigma_z$ are the usual Pauli
matrices, $U_i$ are local single-qubit gate and $\pi/4 \ge h_x \ge
h_y \ge |h_z|$. Benefitting from the nice decomposition
($\ref{decomposition}$), one need not to reverse the whole setup
because $U^{\dag}$ can be constructed from $U$ directly. Now
suppose we have two unitary operations $U$ and $V$. After applying
the selected gate at most 2 times, we can transform one of them,
e.g., $U$, into $f(U)=e^{i h^U_x \sigma_x \otimes \sigma_x}$. If
$f(V) \neq e^{i h^V_x \sigma_x \otimes \sigma_x}$ for some $h^V_x
\in \mathbb{R}$, we can employ the manipulation
$g(\cdot)=Af(\cdot)A^{\dag}f(\cdot)$($A=\sigma_y \otimes I,
\sigma_z \otimes I, I \otimes \sigma_y, \mbox{ or } I \otimes
\sigma_z$), where $A$ can be selected to meet the requirement,
i.e., to reduce the original $U$ and $V$ to $I$ and $g(V)$
respectively. Similarly, by running $g(V)$ at most 4 times, we can
then obtain two local unitary transformations $U'$ and $V'$. One
can easily check that by choosing suitable single-qubit gates,
$U'$ and $V'$ can always be different. Therefore, after repeating
the selected gate at most 20 times, we reduce the original problem
to the discrimination of two local gates, which can be perfectly
implemented with the method we described in the former context.

The same question can also be investigated in multi-partite case.
To answer this problem, we should introduce the generalized
version of the primitive gates. We call $U_{12\ldots N}$ is
$\{[i_s,\ldots,i_e],\ldots, [j_s,\ldots,j_e],\ldots\}$-primitive
if $U_{12\ldots N}$ together with all single-qudit gates can
generate the group $\mathcal{U}_{\beta}=U_{i_s\ldots i_e} \otimes
\ldots \otimes U_{i_s\ldots i_e}$. Similarly, if
$\mathcal{U}_{\beta}=U(d^N)$, then $U_{12\ldots N}$ is
imprimitive. Following the same routine in \cite{exactuniverse},
we can obtain that a $\{[i_s,\ldots,i_e],\ldots,
[j_s,\ldots,j_e],\ldots\}$-primitive gate can be expressed as
$U_{i_s\ldots i_e} \otimes \ldots \otimes U_{i_s\ldots i_e} \cdot
P_{\{[i_s,\ldots,i_e],\ldots, [j_s,\ldots,j_e],\ldots\}}$, where
$P_{\{[i_s,\ldots,i_e],\ldots, [j_s,\ldots,j_e],\ldots\}}$ is
permutation operator which preserves the structure of the
partition $\{[i_s,\ldots,i_e],\ldots, [j_s,\ldots,j_e],\ldots\}$.
For example, if $U_{12345}$ is $\{[1,2],[3,4],5 \}$-primitive,
then $P_{\{[1,2],[3,4],5 \}}=P_{12,34} \otimes I_5$ or
$I_{12345}$, where $P_{12,34}$ is the swap operation between
Hilbert spaces ${\cal H}_1 \otimes {\cal H}_2$ and ${\cal H}_3
\otimes {\cal H}_4$; if $U_{12345}$ is $\{1,2,3,4,5 \}$-primitive,
then $P_{\{1,2,3,4,5 \}}$ can be any element in the permutation
group $S_5$.

We take 3-partite unitary transformations as an instance.
According to the above discussion, if one of the two 3-partite
unitary transformations $U_{ABC}$ and $V_{ABC}$ is $\{A,B,C
\}$-primitive, then perfect local identification can be realized.
If both of the two selected transformations are imprimitive, then
there exists a sequence $f(U_{ABC})=(H'H)\ldots(H'H)$ with
$H'=U_{ABC}HU_{ABC}^{\dag}$, such that $f(U)=e^{i L_{12}^A \otimes
L_{12}^B \otimes L_{12}^C}$. Following the discussion of bipartite
case, we conclude that $U_{ABC}$ and $V_{ABC}$ can be locally
discriminated. Finally, if $U_{ABC}$ is $\{[A,B],C \}$-primitive
with $V_{ABC}$ being $\{A,[B,C] \}$-primitive, then there exists a
circuit such that $f(U_{ABC})=(U_{A_1}\otimes P_{B_1} +
U_{A_2}\otimes P_{B_2} )\otimes U_C$ and $f(V_{ABC})=V_A \otimes
V_{BC}$, where $(U_{A_1}\otimes P_{B_1} + U_{A_2}\otimes P_{B_2}
)$ is some control-unitary transformation. Since $U_{A_1} \neq
V_A$ or $U_{A_2} \neq V_A$, by choosing suitable input state, the
original problem cane be reduced to the discrimination of two
different local unitary manipulations, hence can be realized
perfectly.

The above discussion can be extended to $N$-partite case, and we
have that it is always possible to discriminate two unitary
operations locally, although in general, we need to run the
unknown operation many times. Interestingly, unlike the previous
results for quantum states, where ``the hidden entanglement" plays
a very important role, it seems that the nonlocality of unitary
transformations doesnot affect the distinguishability much (in
this work, it only changes the total run times $N$). We can also
generalize this result to the case of $M$ unitary transformations.
To discriminate the unknown operation from others, we should
perform $M-1$ tests; after each test, one of the $M$ operations
can be ruled out. Therefore perfect local identification can be
realized after a finite number of runs of the unknown gate.

One can also consider the same problem for nonunitary
transformations\cite{nonunitary}. For general completely positive
trace preserving operations $\xi_1$ and $\xi_2$, the reverse
transformations donot always exist unless they are unitary.
Moreover, the output states usually are mixed even if we employ a
pure input state, and $\xi_1$, $\xi_2$ may contain common Kraus
operators. To realize perfect identification operation, these
components should have no contribution to the output states. Thus
totally solve this problem seems to be quite complicated.

To summarize, we have shown that besides global operations,
multi-partite unitary transformations can also be discriminated
perfectly with local methods. Nonlocal schemes together with
entangled input states usually can improve the efficiency of the
identification, i.e., we can run the unknown operation less times
to realize perfect discrimination. However, it doesnot affect the
distinguishability of the whole problem. In principle, by running
the secretly chosen operations a finite number of times, we can
also realize perfect identification under LOCC. From the practical
viewpoint, one need to provide an optimal methods to implement the
discrimination operations. Our investigation indicates that this
question has a close relation to the exact universality of unitary
evolution and the optimal quantum circuit in $d$-level
system\cite{construction}.

The authors thank R. Duan for helpful comments and suggestions and
drawing our attention to their closely related works
\cite{duan}.This work was funded by the National Fundamental
Research Program (2006CB921900) , the National Natural Science
Foundation of China (Grant No. 10674127), the Innovation Funds
from the Chinese Academy of Sciences, and Program for New Century
Excellent Talents in University.

\renewcommand{\theequation}{A-\arabic{equation}}
\setcounter{equation}{0}  
\section*{APPENDIX}  

We now present a simple proof about the exact universality of
$N$-partite unitary transformations. The method used here are
mainly based on ref. \cite{exactuniverse}. First we introduce the
following lemma.

\begin{lemma}
Let $G$ be a compact Lie group. If $H_1$, $\ldots$, $H_k$ are
closed connected subgroups and they generate a dense subgroup of
$G$, then in fact they generate $G$.
\end{lemma}

Suppose $U$ is a $N$-partite unitary map, and we also use $H$ to
denote all $1$-qudit gates $V_1\otimes \ldots \otimes V_N$. We
introduce the subgroup $H_1 = U H U^{-1}$. Now consider the
$n$-fold products $\Sigma^n= \Sigma \ldots \Sigma $ with $\Sigma=
H_1 H$. One can find that when $n \rightarrow \infty$,
$\Sigma^{\infty}$ is a subgroup of all $N$-partite unitary
transformations $U(d^N)$, hence we have $H \subseteq
\Sigma^{\infty} \subseteq U(d^N)$.

Assume $h$, $r$, $g$ are the corresponding Lie algebras of the
group $H$, $\Sigma^{\infty}$, $U(d^N)$ separately. Consider the
representation of $K= SU(d)\otimes \ldots \otimes SU(d)$ on the
Lie algebra $g$
\[
 \pi^{S_1,\ldots, S_N}(\xi)=(S_1\otimes \ldots
\otimes S_N) \xi (S_1\otimes \ldots \otimes S_N)^{\dag},
\hspace{5mm} \xi \in g .
\]
Since $K$ is a compact Lie group, $\pi$ can be decomposed as a
direct sum of irreducible representations of $K$. Therefore, we
obtain the following decomposation of $g$
\begin{eqnarray} \label{A1}
g=\bigoplus_{j=0}^N \bigoplus_{k=1}^{n_j} i^{N+1-j}
P_{[\alpha^{j,k}_1, \ldots \alpha^{j,k}_j]}
\end{eqnarray}
with
\begin{eqnarray}
P_0 &=& \mathbb{R} I \otimes \ldots \otimes I, \\
P_{[\alpha^{j}_1, \ldots \alpha^{j}_j]}&=& I_1 \otimes \ldots
\otimes su(d)_{\alpha^{j}_1} \otimes \ldots \otimes
su(d)_{\alpha^{j}_j} \otimes \ldots,
\end{eqnarray}
where $i^2=-1$, $su(d)$ is the Lie algebra of $SU(d)$, and
$\alpha^{j}_{j'}$ are indices selected from the set $\{ 1, \ldots,
N \}$.

Similarly, because $H \subseteq \Sigma^{\infty}$, $r$ can also be
decomposed into the direct sum of a finite number of terms on the
right-hand-side of Eq. ($\ref{A1}$)
\begin{eqnarray}
r=\bigoplus_{j=0}^{n\le N }\bigoplus_{k=1}^{n_j} c_{jk}
P_{[\alpha^{j,k}_1, \ldots \alpha^{j,k}_j]},  \hspace{5mm} and
\hspace{5mm} c_{jk} \in \{\pm 1, \pm i \}.
\end{eqnarray}

We call two indices $\alpha^{j,k}_{L}$ and $\alpha^{j',k'}_{L'}$
to be connected if there exists a subset $C=[\alpha^{j,k}_1,
\ldots \alpha^{j,k}_j]$ such that $\alpha^{j',k'}_{L'} \in C$ and
$\alpha^{j',k'}_{L'} \in C$. Thus the connectedness of indices
lead to the following decomposition of $\{ 1, \ldots, N \}$
\begin{eqnarray}
[\ldots,\alpha^{j_1,k_1}_{L_1}, \ldots ] \oplus
[\ldots,\alpha^{j_2,k_2}_{L_2}, \ldots ] \oplus \ldots.
\end{eqnarray}

On the other hand, since $r$ is a Lie algebra, one can immediately
obtained that $r$ is the Lie algebra of the compact Lie group
$\mathcal{U}_{\alpha}=U_{[\ldots,\alpha^{j_1,k_1}_{L_1}, \ldots
]}\otimes U_{[\ldots,\alpha^{j_2,k_2}_{L_2}, \ldots ]} \otimes
\ldots$. According to Lemma 2, we obtain that
$\Sigma^{\infty}=\mathcal{U}_{\alpha}$, hence there exist some $p$
such that $\Sigma^p=\mathcal{U}_{\alpha}$.

After we have obtained the group $\mathcal{U}_{\alpha}$, we can
now define the new $n$-fold products as $ \Sigma_1^n=\Sigma_1
\ldots \Sigma_1$ with $\Sigma_1 =
(U\mathcal{U}_{\alpha}U^{\dag})\cdot \mathcal{U}_{\alpha}$. Repeat
the above discussions, we have that $U$ together with all
$1$-qudit gates can generate the following unitary group
\begin{eqnarray}
\mathcal{U}_{\beta}= U_{[\ldots,\beta^{j_1,k_1}_{L_1}, \ldots
]}\otimes U_{[\ldots,\beta^{j_2,k_2}_{L_2}, \ldots ]} \otimes
\ldots
\end{eqnarray}
with $U \mathcal{U}_{\beta} U^{\dag}= \mathcal{U}_{\beta}$.
Therefore, $U$ is $\{[\ldots,\beta^{j_1,k_1}_{L_1}, \ldots ],
[\ldots,\beta^{j_2,k_2}_{L_2}, \ldots ], \ldots  \}$-primitive.
Moreover, $U$ normalize $\mathcal{U}_{\beta}$. Following the
similar discussion in ref. \cite{exactuniverse}, we have that $U$
can be expressed as $U=U_{\beta} \cdot P_{\beta}$ for some
$U_{\beta} \in \mathcal{U}_{\beta}$, where $P_{\beta}$ is the
corresponding permutation operator of the Hilbert spaces ${\cal
H}_{[\ldots,\beta^{j_m,k_m}_{L_m}, \ldots ]}$ which have the same
dimension.

For example, if $U_{12345}$ is $\{[1,2],[3,4],5 \}$-primitive,
then $P_{\{[1,2],[3,4],5 \}}=P_{12,34} \otimes I_5$ or
$I_{12345}$, where $P_{12,34}$ is the swap operation between
Hilbert spaces ${\cal H}_1 \otimes {\cal H}_2$ and ${\cal H}_3
\otimes {\cal H}_4$; if $U_{12345}$ is $\{1,2,3,4,5 \}$-primitive,
then $P_{\{1,2,3,4,5 \}}$ can be any element in the permutation
group $S_5$.


\begin{thebibliography}{99}
\bibitem{discrimination} C.W. Helstron, \emph{Quantum Detection and
Estimation Theory} (Academic Press, New York, 1976).

\bibitem{EPR}  A. Einstein, B. Podolsky, and N. Rosen, Phys. Rev. \textbf{47}, 777 (1935).

\bibitem{unitary} A. Ac\'{i}n, Phys. Rev. Lett. \textbf{87}, 177901 (2001).
\bibitem{unitary2} G.M. D'Ariano, P. Lo Presti, and M.G. A. Paris, Phys. Rev.
Lett. \textbf{87}, 270404 (2001).

\bibitem{gates} D.P. DiVincenzo, Phys. Rev. A \textbf{51}, 1015 (1995); T.
Sleator and H. Weinfurter, Phys. Rev. Lett. \textbf{74}, 4087
(1995); A. Barenco, Proc. R. Soc. London, Ser. A \textbf{449}, 678
(1995); S. Lloyd, Phys. Rev. Lett. \textbf{75}, 346 (1995); D.
Deutsch \emph{et al.}, Proc. R. Soc. London, Ser. A 449, 669
(1995); A. Barenco et al., Phys. Rev. A \textbf{52}, 3457 (1995);
G. Vidal and C.M. Dawson, Phys. Rev. A \textbf{69}, 010301(R)
(2004).

\bibitem{nonlocal1} A. Jamiokowski, Rep. Math.
Phys. \textbf{3}, 275 (1972); J.I. Cirac, W. D\"{u}r, B. Kraus,
and M. Lewenstein, Phys. Rev. Lett. \textbf{86}, 544 (2001).
\bibitem{nonlocal2}W. D\"{u}r, G. Vidal, and J.I. Cirac, Phys.
Rev. Lett. \textbf{89}, 057901 (2002).

\bibitem{locc1} J. Walgate, A.J. Short, L. Hardy and V. Vedral, Phys. Rev. Lett. \textbf{85}, 4972
(2000).
\bibitem{locc2} C.H. Bennett \emph{et al.}, Phys. Rev. A. \textbf{59}, 1070 (1999).

\bibitem{exactuniverse} J.-L. Brylinski and R. Brylinski, \emph{Mathematics of Quantum
Computation}, edited by R. Brylinski and G. Chen (CRC Press, Boca
Raton, 2002) or see quant-ph/0108062.

\bibitem{duanprivite} R. Duan, private communication.

\bibitem{exact-example} M.J. Bremner \emph{et al.}, Phys. Rev. Lett. \textbf{89}, 247902
(2002).

\bibitem{two-qubit} N. Khaneja, R. Brockett, and S.J. Glaser,
Phys. Rev. A \textbf{63}, 032308 (2001); B. Kraus and J.I. Cirac,
Phys. Rev. A \textbf{63}, 062309 (2001).

\bibitem{nonunitary} A. Chefles, Phys. Rev. A \textbf{72}, 042332 (2005);
G.M. D'Ariano, M.F. Sacchi, and J. Kahn, Phys. Rev. A \textbf{72},
052302 (2005); G.M. D'Ariano, P. Mataloni, and M.F. Sacchi, Phys.
Rev. A \textbf{71}, 062337 (2005); Z. Ji, Y. Feng, R. Duan, and M.
Ying, Phys. Rev. Lett. \textbf{96}, 200401 (2006).


\bibitem{construction} A. Muthukrishnan and C.R.Stroud Jr., Phys. Rev. A \textbf{62},
052309 (2000); J.J. Vartiainen, M. M\"{o}tt\"{o}nen, and M.M.
Salomaa, Phys. Rev. Lett. \textbf{92}, 177902 (2004); S.S.
Bullock, D.P. O¡¯Leary, and G.K. Brennen, Phys. Rev. Lett.
\textbf{94}, 230502 (2005); M.A. Nielsen, M.R. Dowling, M. Gu, and
A.C. Doherty, Science \textbf{311}, 1133 (2006).

\bibitem{duan}  R. Duan, Y. Feng, and M. Ying,
Phys. Rev. Lett. 98, 100503 (2007); R. Duan, Y. Feng, M. Ying,
eprint arXiv:0705.1424.
\end{thebibliography}
\end{document}